\documentclass[preprint,showpacs,preprintnumbers,amsmath,amssymb]{revtex4}
\usepackage{graphicx}

\begin{document}
\title{Structural, transport, magnetic properties and Raman spectroscopy of orthorhombic
Y$_{1-x}$Ca$_x$MnO$_3$ ($0 \leq x \leq 0.5$)}
\author{M. N. Iliev$^1$, B.~Lorenz$^1$, A.~P.~Litvinchuk$^1$, Y.-Q.~Wang$^1$,
Y.~Y.~Sun$^1$, C. W. Chu$^{1,2,3}$} \affiliation{$^1$Texas Center
for Superconductivity and Advanced
   Materials and Department of Physics, University of Houston,
   Houston, Texas 77204-5002\\
   $^2$ Lawrence Berkeley National Laboratory, Cyclotron Road, Berkeley, CA 94720\\
$^3$ Hong Kong University of Science and Technology, Hong Kong,
China}
\date{\today}
\begin{abstract}
Orthorhombic Y$_{1-x}$Ca$_x$MnO$_3$ ($0 \leq x \leq 0.5$) was
prepared under high pressure and the variations with $x$ of its
structural, magnetic, electrical properties and the polarized
Raman spectra were investigated. The lattice parameters change
systematically with $x$. Although there are strong indications for
increasing disorder above $x = 0.20$, the average structure
remains orthorhombic in the whole substitutional range. Ca doping
increases conductivity, but temperature dependence of resistivity
$\rho$(T) remains semiconducting for all $x$. The average magnetic
exchange interaction changes from antiferromagnetic for $x < 0.08$
to ferromagnetic for $x > 0.08$. The evolution with $x$ of the
Raman spectra provides evidence for increasingly disordered oxygen
sublattice at $x \geq 0.10$, presumably due to quasistatic and/or
dynamical Jahn-Teller distortions.
\end{abstract}
 \maketitle
\section{Introduction}
At atmospheric pressure Y$_{1-x}$Ca$_x$MnO$_3$ ($0 \leq x <
0.25$), similarly to the rare-earth manganites $R$MnO$_3$, for $R$
with smaller ionic radius ($R$~=~Ho, Er, Tm, Yb, Lu) crystallizes
in the hexagonal $P6_3cm$ structure whereas the orthorhombic,
distorted perovskite structure is the stable phase for
Y$_{1-x}$Ca$_x$MnO$_3$ with $x \geq 0.25$.\cite{moure1} Upon
annealing under high pressure, some hexagonal manganites, such as
YMnO$_3$ and HoMnO$_3$, can be converted to their orthorhombic
$Pnma$ phase.\cite{waintal1} The Ca- and Sr-doped orthorhombic
rare-earth manganites have been a subject of intensive studies
since the phenomenon "colossal magnetoresistance"(CMR) has been
rediscovered.\cite{coey1} To our knowledge, there are however no
reports on the synthesis and properties of doped yttrium-based
orthorhombic manganites for $x<0.25$. The Y$_{1-x}$Ca$_x$MnO$_3$
orthorhombic system is of particular interest as it could be
compared to the model La$_{1-x}$Ca$_x$MnO$_3$ system, where Ca
doping and conversion of part of the Jahn-Teller Mn$^{3+}$ ions in
non-Jahn-Teller Mn$^{4+}$ ions results in a complex magnetic phase
diagram and strong interplay between structural, magnetic and
transport properties.

In this work we report the synthesis of orthorhombic series
Y$_{1-x}$Ca$_x$MnO$_3$ and first results on the variations with
$x$ of structural parameters, resistivity, magnetic properties and
Raman spectra over the doping range $0 \leq x \leq 0.5$. The
lattice parameters change systematically with $x$. Although there
are strong indications for increasing disorder above $x = 0.20$,
the average structure remains orthorhombic in the whole
substitutional range. Ca doping increases conductivity, but the
temperature dependence of the resistivity $\rho$(T) remains
semiconducting for all $x$. It was found, however, that the
average magnetic exchange interaction changes from
antiferromagnetic for $x < 0.08$ to ferromagnetic for $x > 0.08$.
The evolution with $x$ of the Raman spectra provide strong
indications for increasingly disordered oxygen sublattice at $x
\geq 0.10$, presumably due to quasistatic and/or dynamical
Jahn-Teller distortions.

\section{Samples and Experiment}

As a first step in samples preparation, ceramic pellets of
hexagonal Y$_{1-x}$Ca$_x$MnO$_3$ ($x$ = 0.00, 0.03, 0.06, 0.10 and
0.20) and orthorhombic Y$_{1-x}$Ca$_x$MnO$_3$ ($x$ = 0.30 and
0.50) were synthesized by solid state reaction. The prescribed
amounts of Y$_2$O$_3$, CaCO$_3$, and Mn$_2$O$_3$ were mixed and
preheated at 900 to 1000$^\circ$ C in oxygen for 16 hours followed
by sintering at 1140 to 1170$^\circ$C for one day under oxygen
atmosphere. The hexagonal samples were resintered in a high
pressure furnace under 35 kbar at 1015 to 1030~$^\circ$C for 5~h.
Under these conditions the hexagonal phase was completely
transformed into the metastable orthorhombic structure. The
x-ray-diffraction (XRD) pattern were collected at room temperature
using a Rigaku DMaxIII/B x-ray diffractometer.

Magnetization measurements were conducted employing the MPMS SQUID
magnetometer (Quantum Design) at an external field of 500~Oe. The
dc resistivity was measured by the standard four-probe method
using a Keithley~220 current source and a Keithley~182
nanovoltmeter. The Raman spectra were obtained at room temperature
using a HR640 spectrometer equipped with microscope, notch filters
and liquid-nitrogen-cooled CCD detector. 514.5~nm and 632.8 laser
lines, focused with an $\times 100$ objective on the sample's
surface in a spot of (2-3~$\mu$m) diameter, were used for
excitation. For the most samples the size of the the microcrystals
constituting the pellet was larger than the laser spot, which made
it possible to select microcrystals with proper orientation of
crystallographic axes with respect to incident polarization and
obtain polarized spectra in nearly exact scattering
configurations.

\section{Results and Discussion}

\subsection{Structural, Electric and Magnetic Measurements}

The x-ray spectra of Y$_{1-x}$Ca$_x$MnO$_3$ are shown in Fig.1 for
$x=0$ to $x=0.5$. All reflections can be indexed within the
orthorhombic symmetry, space group $Pnma$. The absence of impurity
lines proves the high phase purity and the success of the
high-pressure synthesis in stabilizing the $Pnma$ structure for
the Ca content between 0 and 0.5. A remarkable broadening of the
x-ray lines is observed above $x=0.2$ that indicates an increased
disorder of the cations at higher substitution, $x$. The lattice
constants change systematically with $x$, as shown in Fig.2. The
largest change is the decrease of $a$ by up to 7\% at $x=0.5$. In
contrast, the $c$-axis is almost not affected by the Ca doping.
The $b$-axis shows a distinct, but moderate increase from $x=0$ to
$x=0.5$, the total change being no more than 1.5\%. The systematic
change of the lattice constants with $x$ and the absence of
impurity lines in the x-ray spectra unambiguously prove that the
Ca ions in fact replace the Y ions in the perovskite-like
structure of Y$_{1-x}$Ca$_x$MnO$_3$. Our values of the lattice
parameters are in good agreement with data of recent publications
for $x=0$, $x=0.25$, $x=0.3$, and $x=0.5$ (included in Fig.2 as
symbols).\cite{Brinks,Pollert,Mathieu, Laberty} Recent neutron
powder diffraction (NPD) experiments\cite{munoz1} estimated a
considerably longer $a$-axis (5.97~\AA \ as compared with
5.857~\AA \ of the present investigation). The discrepancy is
probably due to the excess oxygen reported for the NPD samples.

\begin{figure}
\includegraphics[width=3.4in]{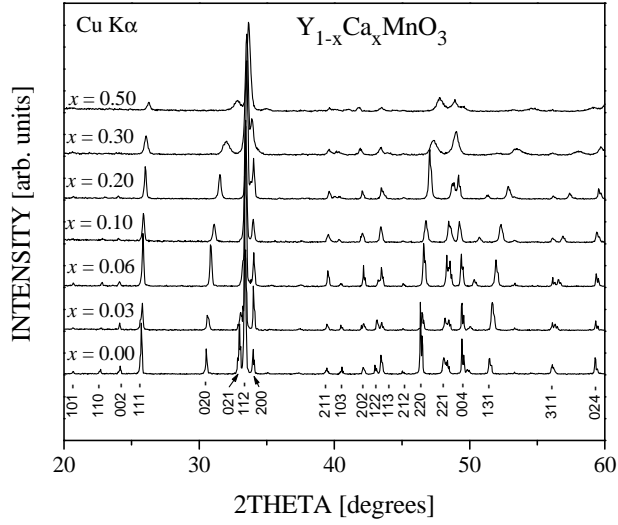}
\caption{X-ray diffraction pattern of Y$_{1-x}$Ca$_x$MnO$_3$ ($0
\leq x \leq 0.5$.)}
\end{figure}

\begin{figure}
\includegraphics[width=2.2in]{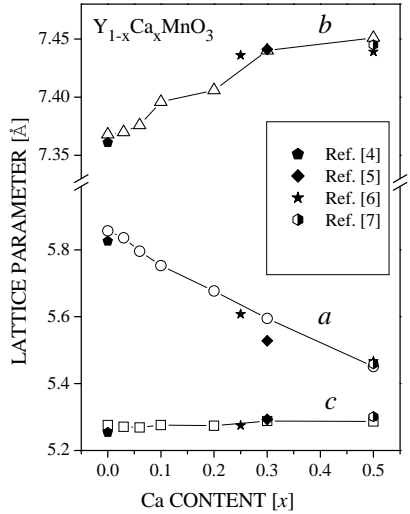}
\caption{Variation of the lattice parameters of
Y$_{1-x}$Ca$_x$MnO$_3$ with $x$}
\end{figure}

The doping of orthorhombic YMnO$_3$ with Ca introduces charge
carriers by removing one electron per Ca from the Mn$^{3+}$~ions.
The resulting Mn$^{4+}$ "carries" one hole that can hop among the
Mn-ions via the hybridization of Mn~$d$-states with the oxygen
$p$-states. The conductivity of the doped compound should increase
with $x$. In fact, the room temperature conductivity of
ortho-YMnO$_3$ increases by a factor of about 300 between $x=0$
and $x=0.5$ (see the inset of Fig. 3). The characteristic
temperature dependence of the resistivity, $\rho$(T), is
semiconducting for all $x<0.5$. Fig.3 (main panel) shows two
typical examples of $\rho$(T) for $x=0.1$ and $x=0.3$.
\begin{figure}
\includegraphics[width=3in]{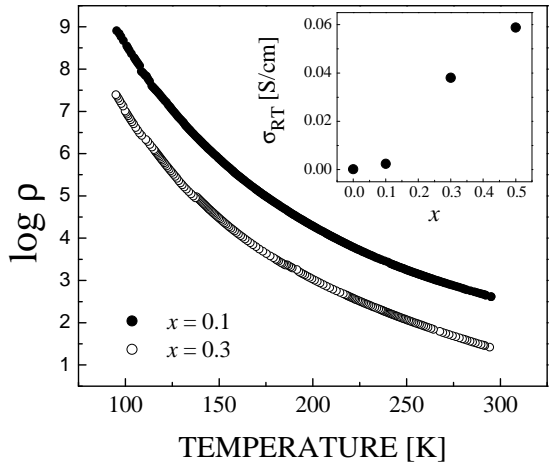}
\caption{Resistivity of Y$_{1-x}$Ca$_x$MnO$_3$ as function of
temperature (for $x=0.1$ and $x=0.3$). Inset: Room temperature
conductivity as function of $x$.}
\end{figure}

The magnetic properties of the series Y$_{1-x}$Ca$_x$MnO$_3$ are
summarized in Fig.4. For $x=0$ the inverse magnetic susceptibility
exhibits a Curie-Weiss-like linear temperature dependence at T~$>
80$~K and a sharp anomaly at the antiferromagnetic transition
temperature, T$_N =42.5$~K. From the high-temperature data the
paramagnetic Curie temperature and the effective magnetic moment
per Mn~ion are estimated as $\Theta = -54.7$~K and $\mu_{eff} =
5.05~\mu_B$, respectively. The effective moment is close to the
value of 4.9~$\mu_B$ for the free Mn$^{3+}$ ion (spin only, S=2).
The values for $\Theta$ and $\mu_{eff}$ are in good agreement with
those reported in Ref.\cite{wood1}(-67 K, 4.98$\mu_B$), but
deviate significantly from the recent results (-26 K, 3.69$\mu_B$)
of Ref.\cite{munoz1}. We attribute the latter discrepancy to the
presence of a significant amount of hexagonal YMnO$_3$ and excess
oxygen in the samples of Ref.\cite{munoz1}. The high-pressure
synthesized specimens used in the present investigation are of
high phase purity. The clear decrease of the magnetic
susceptibility below T$_N$ is a further indication that the
composition of the compound is close to stoichiometric. Deviations
from stoichiometry (e.g. excess oxygen) will result in the
presence of Mn$^{4+}$ ions with a paramagnetic contribution to the
susceptibility that can easily be detected below T$_N$.

The doping of ortho-YMnO$_3$ with Ca$^{2+}$ results in a change of
magnetic properties. Up to $x=0.3$ the high-temperature
susceptibility still exhibits Curie-Weiss behavior (Fig. 4). With
increasing $x$ the effective moment changes very little but
$\Theta$ increases rapidly from -55~K ($x=0$) to +63~K at $x=0.3$.
Fig. 5 shows $\Theta$  as function of $x$. The paramagnetic Curie
temperature crosses zero close to 8\% of Ca doping. The change of
sign of $\Theta$  indicates a change of the average magnetic
exchange interaction from antiferromagnetic (AFM, $x<0.08$) to
ferromagnetic (FM, $x>0.08$), as illustrated in Fig. 5. This
change is a consequence of the presence of Mn$^{4+}$ ions for
$x>0$. For $x=0$ the magnetic exchange of two neighboring
Mn$^{3+}$ is mediated by the oxygen ions between them and is
ascribed to the superexchange mechanism with a resulting AFM
coupling of the moments. With Ca doping Mn$^{4+}$ ions are
created. These Mn$^{4+}$ ions do not participate in the
superexchange interaction but open the double exchange interaction
channel whenever they are next to an Mn$^{3+}$. The double
exchange is ferromagnetic in its nature because of a gain in
kinetic energy (the hole at the Mn$^{4+}$ can hop to the
neighboring Mn$^{3+}$) if the magnetic moments of both Mn-ions are
parallel. This explains the gradual crossover of the paramagnetic
Curie temperature from negative to positive values with increasing
$x$ as estimated from the high-T susceptibility data. Note that
$\Theta$ obtained from the data shown in Fig.5 is an effective
quantity, characteristic for the average response to field and
temperature of all Mn~moments, those coupled by AFM superexchange
as well as those coupled by FM double exchange interactions to
their respective neighbors. Our data clearly demonstrate that Ca
doping of orthorhombic YMnO$_3$ is a powerful tool to tune the
spin correlations from AFM superexchange to predominatly FM double
exchange correlations. However, the increasing FM spin coupling
does not introduce ferromagnetic lang range order down to the
lowest temperatures in agreement with recent results for
x=0.3.\cite{Mathieu} The low temperature AFM Neel transition of
ortho-YMnO$_3$ is also affected by the Ca doping. The relative
enhancement of the magnetic susceptibility below T$_N$ for $x>0$
(as compared to $x=0$) is an indication of the presence of
Mn$^{4+}$ ions and their paramagnetic contribution to the spin
susceptibility. The Neel temperature decreases with increasing $x$
due to a gradual reduction of the AFM superexchange correlations
(Fig. 5). The AFM transition is sharp up to $x=0.2$, but it
broadens at higher doping levels. Only data for T$_N$ up to
$x=0.2$ have therefore been included in the figure. The broadening
of the magnetic transition above $x=0.2$ is consistent with the
observed broadening of the x-ray reflections and the Raman bands
(see next Section)) and with the model of short range
ferromagnetism and a spin glass state proposed recently for
x=0.3.\cite{Mathieu}  It indicates the increasing influence of
disorder on the physical properties of
ortho-Y$_{1-x}$Ca$_x$MnO$_3$.
\begin{figure}
\includegraphics[width=3in]{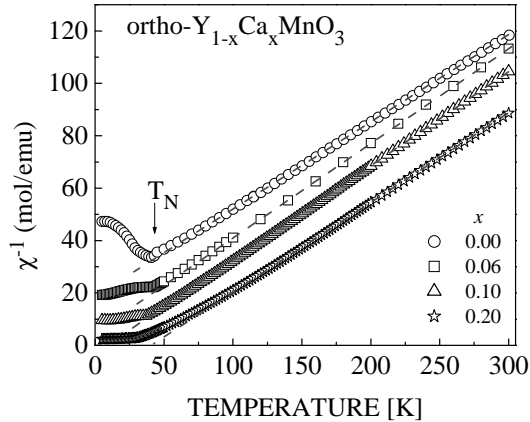}
\caption{Inverse magnetic susceptibility of orthorhombic
Y$_{1-x}$Ca$_x$MnO$_3$ for $x=0$ to $x=0.2$.}
\end{figure}

\begin{figure}
\includegraphics[width=3in]{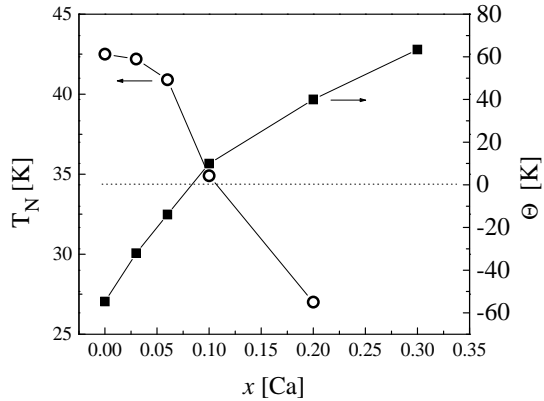}
\caption{Effective paramagnetic Curie temperature of orthorhombic
Y$_{1-x}$Ca$_x$MnO$_3$ estimated from the high-temperature
magnetic susceptibility (filled squares, right scale) and the Neel
temperature for the AFM transition (open circles, left scale).}
\end{figure}

\subsection{Raman Spectra}
The polarized Raman spectra of undoped YMnO$_3$, shown in Fig.6,
are identical to those reported in Ref.\cite{iliev1}, where an
assignment of the lines to particular phonon modes has been done
too. The variations with $x$ of the spectra of
Y$_{1-x}$Ca$_x$MnO$_3$, measured with parallel (HH) and crossed
(HV) scattering configurations, are shown in Fig.7. The HH
configuration is close to $xx$ and the Raman lines correspond to
modes of $A_g$ symmetry. The modes allowed with the HV (nearly
$xz$) configuration are of $B_{2g}$ symmetry.

As it follows from Fig.7, within the doping range $0 \leq x \leq
0.10$ the line widths moderately increase with $x$, but their
position and relative intensity remain practically unchanged. At
$x=0.10$ there is a visible increase in intensity of the high
frequency HH band at 644~cm$^{-1}$ and the HV band at
654~cm$^{-1}$. For $x=0.20$ these bands broaden, increase further
in intensity and shift towards lower wavenumbers along with the
bands originating from the 496~cm$^{-1}$($A_g$) and
617~cm$^{-1}$($B_{2g}$) lines. At $x=0.30$ the overall intensity
of the spectra is reduced. The HV spectrum consists of only two
broad bands centered at 495~cm$^{-1}$ and 624~cm$^{-1}$. Except
for broad bands close to these positions, the corresponding HH
spectrum  contains two additional weak bands at lower wavenumbers.
The evolution of the HV spectrum between $x=0.10$ and $x=0.30$ is
given in more detail in Fig.8. No Raman bands of detectable
intensity are observed for $x=0.5$.

\begin{figure}
\includegraphics[width=3in]{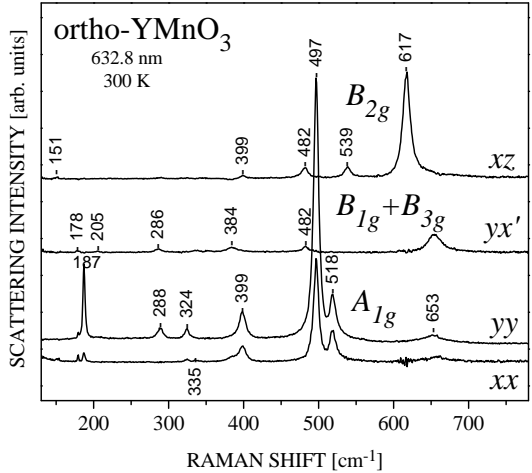}
\caption{Polarized Raman spectra of YMnO$_3$. The assignment of
Raman lines to particular atomic motions is given in
Ref.\cite{iliev1}.}
\end{figure}

\begin{figure}
\includegraphics[width=3in]{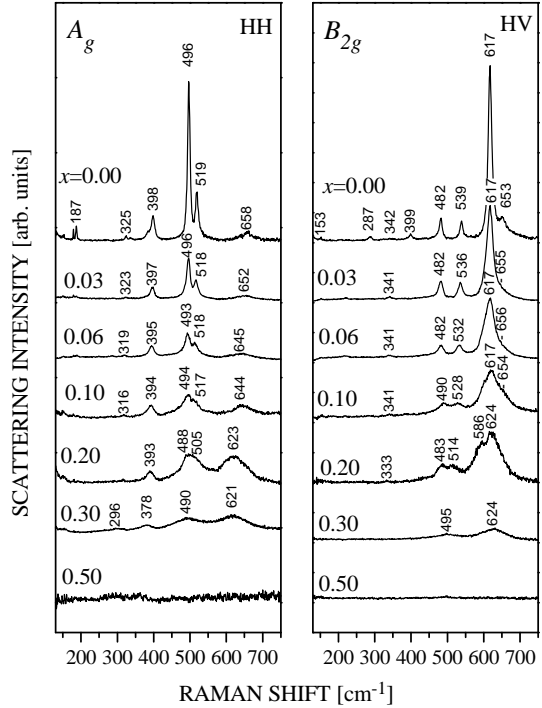}
\caption{Variations with $x$ of the Raman spectra of
Y$_{1-x}$Ca$_x$MnO$_3$. The HH and HV configurations are close,
respectively to $xx$ and $xz$}
\end{figure}

\begin{figure}
\includegraphics[width=3in]{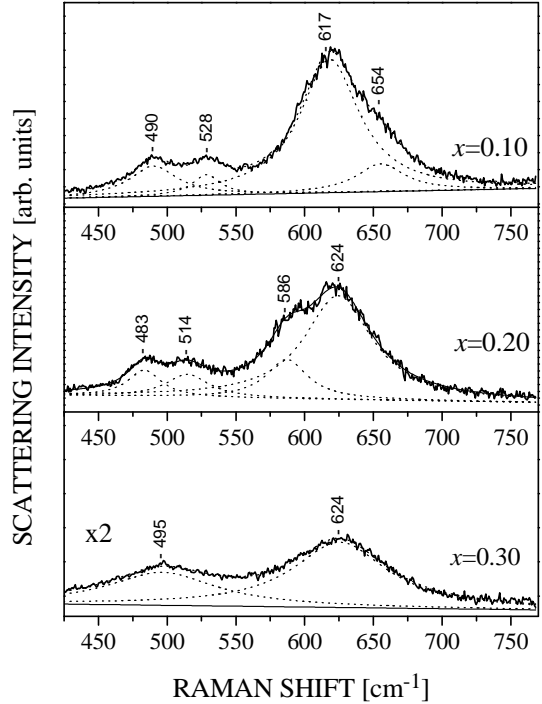}
\caption{Evolution of the HV spectrum between $x=0.10$ and
$x=0.30$.}
\end{figure}

The Raman spectra for $x \geq 0.10$ are similar to those reported
earlier for other $R^{3+}_{1-x}A^{2+}_x$MnO$_3$ ($R=$~rare earth,
$A=$~Ca,Sr,Ba;\ $x \geq 0.1)$ systems,
\cite{podobedov1,yoon1,granado1,abrashev1,liarokapis1,bjornsson1,iliev2}
where 2 or 3 broad bands are observed at positions close to those
of the strongest lines for nominally undoped RMnO$_3$. On the
basis of this closeness, the broad bands have usually been
assigned to the corresponding Raman allowed modes in the parent
$Pnma$ structure.

An alternative explanation for the broad band origin has recently
been proposed by Iliev et al.\cite{iliev3}. It has been argued
that at higher doping levels the spectral profiles reflect smeared
partial phonon density-of-states (PDOS) related to oxygen
vibrations rather than broadened $\Gamma$-point Raman allowed
phonon modes. Indeed, the coexistence in doped manganites of
Jahn-Teller-distorted (Mn$^{3+}$O$_6$) and undistorted
(Mn$^{4+}$O$_6$) octahedra and the Mn$^{3+} \leftrightarrows$
Mn$^{4+}$ charge transfer results in strong quasistatic or/and
dynamical Jahn-Teller disorder of the oxygen sublattice. The loss
of translational symmetry activates otherwise Raman forbidden
oxygen vibrations, corresponding to the off-center phonon modes in
ordered parent compound. In support of this model are: (i) the
good correspondence between experimental Raman profiles and
calculated smeared PDOS; (ii) the consistent explanation of the
similarities between the Raman spectra of manganites with
different average structure (orthorhombic or rhombohedral); (iii)
the strong reduction or disappearance of the broad bands below the
insulator-to-metal transition in the CMR materials.

Our results support the assumption for dominant role of
Jahn-Teller-disorder-induced bands in the Raman spectra of heavily
doped orthorhombic manganites. Indeed, the Raman spectrum of
undoped YMnO$_3$ is well understood and the $A_g$ line at
519~cm$^{-1}$ and the $B_{2g}$ line at 617~cm$^{-1}$ correspond to
the highest $A_g$ and $B_{2g}$ modes, respectively.\cite{iliev1}
Therefore, the relatively weak band at 653-658~cm$^{-1}$, observed
in the HH and HV spectra of YMnO$_3$, cannot be a proper Raman
mode for the $Pnma$ structure, but is rather due to contributions
from zone-boundary phonons. This is consistent with lattice
dynamical calculations, which predict for both YMnO$_3$ and
LaMnO$_3$ the strongest PDOS peak close to this
frequency.\cite{iliev3}  This structure and another one near
490~cm$^{-1}$, where a PDOS maximum is also predicted, grow in
relative weight with increasing $x$ and become dominant at
$x=0.30$, while the lines related to $\Gamma$-point Raman modes,
involving mainly oxygen motions are diminishing. This is exactly
what one expects upon loss of translation symmetry with increasing
oxygen disorder. It is worth noting that the cationic sublattice
also exhibits increasing disorder at higher $x$, indicated by
noticeable broadening of the x-ray diffraction patterns for
$x=0.30$ and $x=0.50$ (Fig.1). The x-ray patterns gain their
intensities mainly from the Y, Ca and Mn atoms and to a lesser
extent from the light oxygen atoms, which justifies such
conclusion.

\section{Conclusions}
We have successfully prepared orthorhombic Y$_{1-x}$Ca$_x$MnO$_3$
in a wide doping range between $x=0$ and $x=0.5$ using
high-pressure synthesis. Between $x=0$ and $x=0.25$ the
orthorhombic structure was stabilized as a metastable phase. The
lattice parameters change gradually with $x$ over the whole doping
range. The Ca doping increases the conductivity, but its
temperature dependence remains semiconducting for all $x$. The
magnetic susceptibility for $x \leq 0.3$ exhibits the typical
paramagnetic Curie dependence at high temperatures ($>80$~K). The
paramagnetic Curie temperature is negative for small $x$, crosses
zero at $x \approx 0.08$ and becomes positive at larger $x$. This
is interpreted as a gradual change of the magnetic correlations
from antiferromagnetic superexchange ($x<0.08$) to ferromagnetic
double exchange interactions ($x>0.08$) due to the replacement of
Mn$^{3+}$ with Mn$^{4+}$ with increasing Ca doping. The evolution
with $x$ of the Raman spectra provide strong indications for
increasing disorder of the oxygen sublattice for $x \geq 0.10$,
presumably due to quasistatic and/or dynamic Jahn-Teller
distortions. As a consequence of the loss of translational
symmetry and activation of otherwise forbidden vibrations, the
relatively broad bands in the Raman spectra for high substitution
levels  reflect rather the phonon density of states than the Raman
allowed zone center phonons.

\acknowledgments This work is supported in part by the state of
Texas through the Texas Center for Superconductivity and Advanced
Materials, by NSF grant no. DMR-9804325, the T.L.L. Temple
Foundation, the J. J. and R. Moores Endowment, and at LBNL by the
Director, Office of Energy Research, Office of Basic Energy
Sciences, Division of Materials Sciences of the US Department of
Energy under contract no. DE-AC03-76SF00098.

\end{document}